\documentclass[a4paper,prl,showpacs,superscriptaddress,nofootinbib,twocolumn]{revtex4-1}
\usepackage[utf8]{inputenc}
\usepackage[T1]{fontenc}
\usepackage[UKenglish]{babel}
\usepackage[sc,osf]{mathpazo}
\usepackage[scaled=0.86]{berasans}
\usepackage[scaled=1.03]{inconsolata}
\usepackage[colorlinks]{hyperref}
\usepackage{graphicx}
\usepackage{amsmath,amssymb,amsthm}
\usepackage{xspace}
\usepackage{mathtools}

\let\originalleft\left
\let\originalright\right
\renewcommand{\left}{\mathopen{}\mathclose\bgroup\originalleft}
\renewcommand{\right}{\aftergroup\egroup\originalright}

\newcommand{\ket}[1]{\left| #1 \right\rangle}
\newcommand{\braket}[2]{\left\langle #1 \middle| #2 \right\rangle}

\newcommand{\de}[1]{\left(#1\right)}

\newcommand{\id}{\mathbb{1}}

\newcommand{\eg}{\emph{e.g.}\@\xspace}
\newcommand{\ie}{\emph{i.e.}\@\xspace}
\newcommand{\etal}{\emph{et al.}\@\xspace}

\mathtoolsset{centercolon}

\input{Qcircuit}

\newcommand{\blue}{\ket{\text{b}}}
\newcommand{\red}{\ket{\text{r}}}

\begin{document}

\title{Quantum circuits cannot control unknown operations}

\author{Mateus Araújo}
\author{Adrien Feix}
\author{Fabio Costa}
\author{Časlav Brukner}
\affiliation{Faculty of Physics, University of Vienna, Boltzmanngasse 5, 1090 Vienna, Austria}
\affiliation{Institute of Quantum Optics and Quantum Information (IQOQI), Austrian Academy of Sciences, Boltzmanngasse 3, 1090 Vienna, Austria}


\date{\today}


\begin{abstract}
One of the essential building blocks of classical computer programs is the ``\textbf{if}'' clause, which executes a subroutine depending on the value of a control variable. Similarly, several quantum algorithms rely on applying a unitary operation conditioned on the state of a control system. Here we show that this control cannot be performed by a quantum circuit if the unitary is completely unknown. However, this no-go theorem does not prevent implementing quantum control of unknown unitaries in practice, as any physical implementation of an unknown unitary provides additional information that makes the control possible. We then argue that one should extend the quantum circuit formalism to capture this possibility in a straightforward way. This is done by allowing unknown unitaries to be applied to subspaces and not only to subsystems.
\end{abstract}


\maketitle

	Quantum computation harnesses quantum effects to significantly outperform classical computation in solving specific problems. The most widely used model for quantum computation -- referred to as the \textit{quantum circuit model} -- is formulated in terms of wires, representing quantum systems, which connect boxes, representing unitary operations~\cite{chuang00}. The formalism of quantum circuits is often considered as the standard language for describing quantum algorithms.
	
	The difference between the classical and the quantum models of computation is not only in the computational complexity: quantum information processing differs in fundamental, and often counterintuitive, ways from classical computing. One of the most striking examples is the fact that it is impossible to produce a perfect copy of an unknown quantum state~\cite{wootters82}, whereas copying of classical information is a standard operation for classical computers.
 
	Another standard operation in classical computer programs, usually expressed by an ``\textbf{if}'' clause, is the conditional execution of a part of the program depending on the value of a variable. A typical programming line of this kind has the form ``\textbf{if $x=0$, do $A$}'', where $A$ represents an arbitrary set of commands, \ie, a subroutine. Crucially, the construction of the \textbf{if} clause is independent of the subroutine $A$, allowing the latter to be used as a variable in the program.
 
	The quantum analogue of the ``\textbf{if}'' clause is the control of a unitary operation $U$ depending on the value of a control quantum bit (qubit). This is represented by the transformation
\[ \de{\alpha\ket{0}_{C}+\beta\ket{1}_{C}}\ket{\psi} \mapsto \alpha\ket{0}_{C}\ket{\psi}+\beta\ket{1}_{C}U\ket{\psi}, \]
where the subscript $C$ stands for the control qubit and $\ket{\psi}$ is the initial state of the target system. This control-$U$ gate is a fundamental tool for quantum computation. It is used for example in Kitaev's phase estimation algorithm~\cite{kitaev95}, Shor's factoring~\cite{shor94}, Metropolis sampling~\cite{temme09}, and in the Determinstic Quantum Computing on One Qubit (DQC1) computing model~\cite{knill98}.
 
	The standard strategy to implement this gate in a quantum circuit is to decompose $U$ into elementary gates, for which one knows how to add control~\cite{barenco95}. This approach requires the unitary to be \textit{known}, thus it cannot be used for solving problems in which the unitary itself is a variable. A genuine quantum analogue of the \textbf{if} clause would be an implementation of the control-$U$ gate in which $U$ can be treated as a blackbox. Although it has long been suspected that such a construction is impossible (\eg, it is mentioned \textit{en passant} by Kitaev in Ref.~\cite{kitaev95}), no proof of this fact is known to the best of our knowledge. 

	Classically, the control can be achieved by encoding the operation to be controlled as a bit string in the input, in what is known as the ``von Neumann'' architecture, or a stored-program computer. In the quantum case, however, it is not possible to encode $U$ as an input state, due to the no-programming theorem~\cite{nielsen97}; for this reason, computations in which $U$ is a variable have to be considered as transformations of operations rather than states \cite{gutoski06,chiribella08}.
	
	Here we prove a no-go theorem that states the impossibility of controlling an arbitrary unknown unitary in the quantum circuit model. The formal question is whether a quantum circuit can implement the control-$U$ gate, given as input a single copy of the unknown $d \times d$ gate $U$. Thus we ask whether there exist unitaries $A$ and $B$ such that the following circuit identity is satisfied:
\[
\Qcircuit @C=.5em @R=0em @!R {
\lstick{\ket{0}_{a}} & \multigate{2}{A}  & \qw      & \multigate{2}{B} & \qw &   & \lstick{\ket{0}_{a}} & \gate{W_U}& \qw \\
& \ghost{A} 	    & \qw      & \ghost{B} 	  & \qw & \push{\rule{.3em}{0em}\stackrel{?}{=}\rule{1.3em}{0em}} & & \ctrl{1}  & \qw \\
& \ghost{A}         & \gate{U} & \ghost{B}        & \qw & 							 & & \gate{U}  & \qw 
}
\]
where the topmost line represents an additional $a$-dimensional quantum system (ancilla) and $W_{U}$ is an arbitrary unitary on the ancilla, possibly depending on $U$. Note that the left hand side depicts the most general transformation that a quantum circuit can effect on $U$~\cite{chiribella09b}.

	In order to see that the above identity cannot be satisfied, it is sufficient to notice that, in the \emph{lhs}, substituting $U$ with $e^{i\varphi}U$ does not produce any physical difference, since the two circuits only differ by a global phase $e^{i\varphi}$. In contrast, the same substitution in the \emph{rhs} produces a measurable relative phase.
	
	Since the unitary $U$ in the \emph{lhs} is only defined up to a phase, it is only meaningful to ask whether a circuit can implement the control-$U$ modulo this global phase. To translate this question into an equation, note that the matrix representation of the control-$U$ operation is given by $\id_{d} \oplus U$. Defining $\ket{U}_a:=W_U\ket{0}_a$, the question is whether the identity 
	\begin{equation}\label{eq:controlu}
	B \de{\id_a \otimes \id_2 \otimes U} A\ket{0}_a = \ket{U}_a (\id_d \oplus e^{iu}  U)
	\end{equation}
	holds for some arbitrary phase factor $e^{iu}$, possibly depending on $U$. This is still not possible, due to the non-linearity of the transformation $U \mapsto \id_d \oplus e^{iu} U$. To see this, assume that equation \eqref{eq:controlu} is valid for the qubit unitaries $X$, $Z$, and $H=\alpha X+\beta Z$, where $X$ and $Z$ are Pauli matrices, and $\alpha,\beta$ are real numbers such that $\alpha^2+\beta^2=1$. One has then
	\begin{equation}
	B \big[\id_a \otimes \id_2 \otimes \de{\alpha X+\beta Z} \big] A\ket{0}_a = \ket{H}_a (\id_2 \oplus e^{ih}  H).
	\end{equation}
	Expanding the \emph{lhs} by linearity and using equation \eqref{eq:controlu} again we get
	\begin{multline}
	\alpha \ket{X}_a(\id_2 \oplus e^{ix}  X) + \\ 
	\beta\ket{Z}_a (\id_2 \oplus e^{iz}  Z) =\ket{H}_a (\id_2 \oplus e^{ih}  H).
	\end{multline}
	Taking the inner product with $\ket{H}_{a}$ in the ancilla subsystem gives us the equations
	\begin{subequations}
	\begin{align}\label{eq:firstequation}
	 \alpha \braket{H}{X} + \beta\braket{H}{Z} &= 1, \\
	 \alpha \braket{H}{X} e^{ix}  X + \beta\braket{H}{Z} e^{iz}  Z &= e^{ih}\de{\alpha X + \beta Z}.\label{eq:secondequation}
	\end{align}
	\end{subequations}
	Since $X$ and $Z$ are orthogonal, equation \eqref{eq:secondequation} implies that $\braket{H}{X} =e^{i\de{h-x}}$ and $\braket{H}{Z} =e^{i\de{h-z}}$. Substituting into the equation \eqref{eq:firstequation}
	we get that
	\begin{equation}
	 e^{i\de{h-x}} \de{\alpha + \beta e^{i\de{x-z}}} = 1.
	\end{equation}
	Taking the modulus squared of the equation shows us that $\cos(x-z)=0$. Repeating these calculations for the matrices $\alpha X+\beta Y$ and $\alpha Y+\beta Z$, we get also the conditions $\cos(x-y)=0$ and $\cos(y-z)=0$; but this is a contradiction, since there exist no angles $x$, $y$, and $z$ such that 
	\begin{equation}
	\cos(x-z) = \cos(x-y) = \cos(y-z)=0.
	\end{equation}
 \qed

	This shows that one cannot control an arbitrary unknown unitary in the quantum circuit model. However, it leaves open the possibility of controlling unitaries belonging to known, specific sets. For example, if one eigenvector of $U$ and its eigenvalue are known, there is a circuit that performs the control~\cite{kitaev95}. In a similar vein, if one knows that $U$ belongs to a given set of orthogonal unitaries, it is also possible to control it~\cite{bisio13}.

	Furthermore, it is possible to have quantum control over classical operations on classical inputs. To see this, we restrict the target system $\ket{\psi}$ to be a classical bit string, \ie, to belong to the computational basis, and the unitary to be a classically allowed transformation $U_{\text{cl}}$, \ie, a permutation matrix. Then the following circuit implements the control-$U_{\text{cl}}$:
\[\Qcircuit @C=.5em @R=0.5em @!R {
\lstick{\ket{C}} &\ctrlo{1} & \qw & \ctrl{1} &  \qw\\
\lstick{\ket{\psi}} &\multigate{1}{\mathcal{C}} & \gate{U_{\text{cl}}}
&\multigate{1}{\mathcal{C}} & \qw \\
\lstick{\ket{0}}      &\ghost{\mathcal{C}} & \qw & \ghost{\mathcal{C}} & \qw
}
\]
where $\mathcal{C}$ represents a classical cloning operation (a controlled-NOT for a two-level system). The symbol $\Qcircuit @C=.7em @R=.3em { & \ctrl{1} & \qw\\ & \qwx & }$ means that the operation is applied if the control bit is $\ket{1}$ and $\Qcircuit @C=.7em @R=.3em { & \ctrlo{1} & \qw\\ & \qwx & }$ means that the operation is applied when the control bit is $\ket{0}$. Note that, if cloning an arbitrary quantum state $\ket{\psi}$ were possible, the circuit given above would allow one to control unknown quantum operations. This provides an alternative proof of the no-cloning theorem~\cite{wootters82}.

\begin{figure}[tb]
	\centering
	\includegraphics[width=1.00\columnwidth]{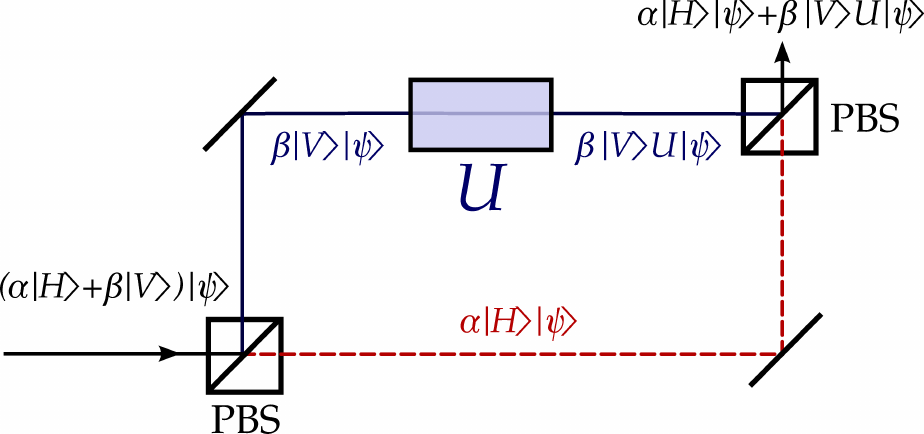}
	\caption{Interferometer that controls a qudit blackbox $U$. Here the control qubit is the polarization of the photon, and $U$ acts on some additional degree of freedom of the same photon. The PBSs are polarizing beam splitters. A photon with polarization $\ket{H}$ takes the lower (red) path, while one with polarization $\ket{V}$ takes the upper (blue) path.}
	\label{fig:polcontrol}
	\end{figure}

	We will now show that, unlike for the no-cloning theorem, our no-go theorem does not prevent quantum control of unknown operations from being performed in practice. In fact, control of blackbox quantum gates has been experimentally demonstrated~\cite{lanyon09,zhou11,zhou13}. In order to illustrate how this is possible, we propose here a simple interferometric setup, depicted in Fig.~\ref{fig:polcontrol}, that exploits a similar idea, but implements the control-$U$ operation in a very direct way.

	Consider a single photon in the state $(\alpha\ket{H}_{C}+\beta\ket{V}_{C})\ket{\psi}$, where $\ket{H}$ and $\ket{V}$ represent horizontal and vertical polarization states of the photon and $\ket{\psi}$ is the state of some other degree of freedom of the same photon (it could be its orbital angular momentum, a qudit of spatial or temporal bins, etc.). Let then $U$ be a unitary gate acting on this additional degree of freedom. It is straightforward to check that the interferometer in Fig.~\ref{fig:polcontrol} applies the transformation
	\begin{equation}\label{eq:interferometro}
(\alpha\ket{H}_{C}+\beta\ket{V}_{C})\ket{\psi} \mapsto \alpha\ket{H}_{C}\ket{\psi}+\beta\ket{V}_{C}U\ket{\psi},
	\end{equation}
for any blackbox unitary $U$.

\begin{figure}[tb]
	\centering
	\includegraphics[width=1.00\columnwidth]{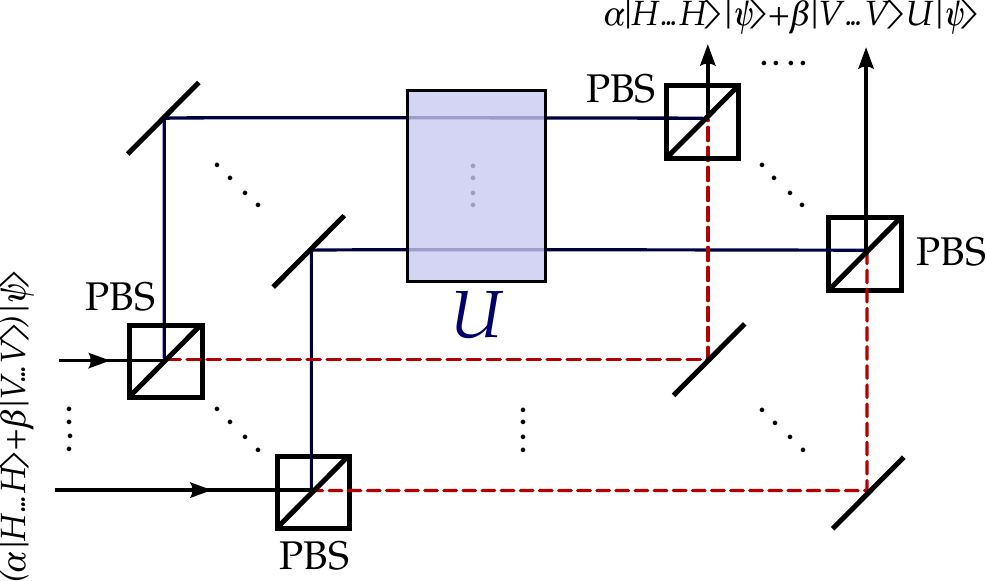}
	\caption{Scalable implementation of an $n$-qubit blackbox $U$. The control qubit is encoded in the polarization state of $n$ photons as $\alpha\ket{H}^{\otimes n} + \beta\ket{V}^{\otimes n}$ (this can be prepared with a linear amount of elementary gates). Each photon goes through a different interferometer and $U$ acts across the upper arms of all interferometers. The total number of PBSs required for this implementation is $2n$.}
	\label{fig:polcontrolscalable}
	\end{figure}

	This interferometer is not scalable, since the whole Hilbert space is encoded in a single photon. One can, however, generalize it to a scalable implementation of a controlled $n$-qubit unitary, in which each qubit is encoded in a different photon, as shown in Fig.~\ref{fig:polcontrolscalable}. 

	How does this interferometric implementation circumvent the no-go theorem we have just proved? The crucial point is the difference between the unitary matrix $U$ that appears in a quantum circuit, and the physical device that implements it  in the interferometer $U_{\text{physical}}$. While $U$ is completely unknown, the \emph{position} of the physical unitary is known, and therefore we know that it acts trivially on modes that do not pass through it. This knowledge is sufficient to control $U$, because when we extend the description of $U$ to include this trivial subspace, we see that it is represented as $U_{physical}=\id_{d} \oplus U$, which is the control-$U$ gate.
	
	To be more explicit, let $\{\ket{0}, \ldots, \ket{d-1}\}$ be a basis spanning the space in which $U$ acts, and $\{\red,\blue\}$ denote the red and blue paths. Then, in the basis $\{\red,\blue\} \otimes \{\ket{0}, \ldots, \ket{d-1}\}$, the physical operation is represented by
	\[ 
	U_{\text{physical}} = \begin{pmatrix} \id_d & 0 \\ 0 & U \end{pmatrix}, 
	\]
	which is exactly the matrix representation of the control-$U$ gate. We stress again that $U$ is still completely unknown, but the physical operation $U_{\text{physical}}$ is not -- some of its eigenvalues are known -- and therefore the theorem does not apply to it.
	
	This knowledge about a subspace in which the physical unitary acts trivially is not a particularity of this photonic implementation, but rather it must be present in any physical implementation of a quantum gate, since every physical operation acts non-trivially only on a limited region of space and time, limited number of electronic or nuclear levels, frequency modes, and so on.
	
	In general, knowing that the physical implementation of a unitary $U$ acts trivially on a $d'$-dimensional subspace allows one to write it as $\id_{d'} \oplus U$; in fact, even a one-dimensional extension $1\oplus U$ allows one to implement a control-$U$ using the scheme of Ref.~\cite{kitaev95}, because one eigenvector and the corresponding eigenvalue of $1\oplus U$ are known.
	
	This is similar to the implicit knowledge that physical operations only act on a limited number of \emph{subsystems}. In the quantum circuit model, this is taken into account by adding additional wires, which corresponds to the map $U \mapsto \id \otimes U$. Therefore, a natural way of accounting for the fact that physical operations act only on restricted \emph{subspaces} would be to consider a generalized circuit model that allows extensions of the form $U \mapsto \id \oplus U$.
	
	Our results have profound implications for quantum algorithms that rely on estimating properties of unknown unitaries. Taking into account the fact that unitaries can be applied to subspaces, it becomes possible to use algorithms such as DQC1 trace estimation~\cite{knill98} and Kitaev's phase estimation~\cite{kitaev95} with unknown unitaries, which would be impossible in the quantum circuit framework. Note that, however, Kitaev's algorithm is not efficient when used on blackboxes, since it must compute the unitaries $U^{2^{k}}$, and this requires an exponential amount of copies of $U$.
		
	Furthermore, the scheme presented could also be used to simplify the implementation of the control-$U$ gate even when the unitary in question is known, since adding control in the traditional way incurs in a constant overhead \cite{barenco95} that, while irrelevant in complexity theory, is usually important for physical implementations. One concrete example would be using the setup of Fig.~\ref{fig:polcontrol} to approximate the Jones polynomial with DQC1~\cite{shor08}.

	In conclusion, we have proved a no-go theorem that shows that an unknown arbitrary unitary cannot be controlled in a quantum circuit, even modulo a global phase. This control is, however, possible for any physical implementation of a unitary transformation. This shows that the language of quantum circuits should be extended in order to capture all information processing possibilities allowed by quantum physics. Other extensions of the quantum circuit formalism have been proposed, in which the wires between gates can be in a superposition~\cite{chiribella09,chiribella12,colnaghi11}. Allowing for such ``superpositions of circuits'' can simplify the implementation of some information processing tasks~\cite{chiribella12,colnaghi11}, or even reduce the computational complexity of some problems~\cite{araujo13b}. It is an intriguing open question whether further extensions are possible~\cite{oreshkov12}. 

	After this work was submitted, we became aware that similar results were obtained independently by Akihito Soeda \cite{soeda13}, and related work was developed by Thompson \etal \cite{thompson13}.
	
\begin{acknowledgments}
We thank Alessandro Bisio, Jan-Åke Larsson, Amir Moqanaki, Daniel Nagaj, and Michal Sedlák for helpful discussions and an anonymous referee for valuable comments. This work was supported by the European Commission Project RAQUEL, the John Templeton Foundation, FQXi, and the Austrian Science Fund (FWF) through CoQuS, SFB FoQuS, and the Individual Project 2462.
\end{acknowledgments}

\bibliographystyle{linksen}
\bibliography{biblio}

\providecommand{\href}[2]{#2}\begingroup\raggedright\begin{thebibliography}{10}

\bibitem{chuang00}
M.~Nielsen and I.~Chuang, {\em Quantum Computation and Quantum Information}.
\newblock Cambridge University Press, 2000.

\bibitem{wootters82}
W.~K. Wootters and W.~H. Zurek, ``A single quantum cannot be cloned,''
  \href{http://dx.doi.org/10.1038/299802a0}{{\em Nature} {\bfseries 299},
  802--803 (1982)}.

\bibitem{kitaev95}
A.~Y. {Kitaev}, ``{Quantum measurements and the Abelian Stabilizer Problem},''
  \href{http://arxiv.org/abs/quant-ph/9511026}{{\ttfamily quant-ph/9511026}}.

\bibitem{shor94}
P.~Shor, ``Algorithms for quantum computation: discrete logarithms and
  factoring,'' \href{http://dx.doi.org/10.1109/SFCS.1994.365700}{{\em
  Foundations of Computer Science, 1994 Proceedings, 35th Annual Symposium on}
  124--134 (1994)}.

\bibitem{temme09}
K.~{Temme}, T.~J. {Osborne}, K.~G. {Vollbrecht}, D.~{Poulin}, and
  F.~{Verstraete}, ``{Quantum Metropolis sampling},''
  \href{http://dx.doi.org/10.1038/nature09770}{{\em Nature} {\bfseries 471},
  87--90 (2011)}, \href{http://arxiv.org/abs/0911.3635}{{\ttfamily
  arXiv:0911.3635 [quant-ph]}}.

\bibitem{knill98}
E.~{Knill} and R.~{Laflamme}, ``{Power of One Bit of Quantum Information},''
  \href{http://dx.doi.org/10.1103/PhysRevLett.81.5672}{{\em Phys. Rev. Lett.}
  {\bfseries 81}, 5672--5675 (1998)},
  \href{http://arxiv.org/abs/quant-ph/9802037}{{\ttfamily quant-ph/9802037}}.

\bibitem{barenco95}
A.~Barenco, C.~H. Bennett, R.~Cleve, D.~P. DiVincenzo, N.~Margolus, P.~Shor,
  T.~Sleator, J.~A. Smolin, and H.~Weinfurter, ``Elementary gates for quantum
  computation,'' \href{http://dx.doi.org/10.1103/PhysRevA.52.3457}{{\em Phys.
  Rev. A} {\bfseries 52}, 3457--3467 (1995)},
  \href{http://arxiv.org/abs/quant-ph/9503016}{{\ttfamily quant-ph/9503016}}.

\bibitem{nielsen97}
M.~A. Nielsen and I.~L. Chuang, ``Programmable Quantum Gate Arrays,''
  \href{http://dx.doi.org/10.1103/PhysRevLett.79.321}{{\em Phys. Rev. Lett.}
  {\bfseries 79}, 321--324 (1997)},
  \href{http://arxiv.org/abs/quant-ph/9703032}{{\ttfamily quant-ph/9703032}}.

\bibitem{gutoski06}
G.~Gutoski and J.~Watrous, ``Toward a general theory of quantum games,'' in
  {\em In Proceedings of 39th ACM STOC}, pp.~565--574.
\newblock 2006.
\newblock \href{http://arxiv.org/abs/quant-ph/0611234}{{\ttfamily
  quant-ph/0611234}}.

\bibitem{chiribella08}
G.~Chiribella, G.~M. D'Ariano, and P.~Perinotti, ``Quantum Circuit
  Architecture,'' \href{http://dx.doi.org/10.1103/PhysRevLett.101.060401}{{\em
  Phys. Rev. Lett.} {\bfseries 101}, 060401 (2008)},
  \href{http://arxiv.org/abs/0712.1325}{{\ttfamily arXiv:0712.1325
  [quant-ph]}}.

\bibitem{chiribella09b}
G.~{Chiribella}, G.~M. {D'Ariano}, and P.~{Perinotti}, ``{Theoretical framework
  for quantum networks},''
  \href{http://dx.doi.org/10.1103/PhysRevA.80.022339}{{\em Phys. Rev.~A}
  {\bfseries 80}, 022339 (2009)},
  \href{http://arxiv.org/abs/0904.4483}{{\ttfamily arXiv:0904.4483
  [quant-ph]}}.

\bibitem{bisio13}
A.~Bisio, P.~Perinotti, and M.~Sedlák (in preparation).

\bibitem{lanyon09}
B.~P. {Lanyon}, M.~{Barbieri}, M.~P. {Almeida}, T.~{Jennewein}, T.~C. {Ralph},
  K.~J. {Resch}, G.~J. {Pryde}, J.~L. {O'Brien}, A.~{Gilchrist}, and A.~G.
  {White}, ``{Simplifying quantum logic using higher-dimensional Hilbert
  spaces},'' \href{http://dx.doi.org/10.1038/nphys1150}{{\em Nat. Phys.}
  {\bfseries 5}, 134--140 (2009)},
  \href{http://arxiv.org/abs/0804.0272}{{\ttfamily arXiv:0804.0272
  [quant-ph]}}.

\bibitem{zhou11}
X.-Q. {Zhou}, T.~C. {Ralph}, P.~{Kalasuwan}, M.~{Zhang}, A.~{Peruzzo}, B.~P.
  {Lanyon}, and J.~L. {O'Brien}, ``Adding control to arbitrary unknown quantum
  operations,'' \href{http://dx.doi.org/10.1038/ncomms1392}{{\em Nat. Commun.}
  {\bfseries 2}, (2011)}, \href{http://arxiv.org/abs/1006.2670}{{\ttfamily
  arXiv:1006.2670 [quant-ph]}}.

\bibitem{zhou13}
X.-Q. {Zhou}, P.~{Kalasuwan}, T.~C. {Ralph}, and J.~L. {O'Brien},
  ``{Calculating unknown eigenvalues with a quantum algorithm},''
  \href{http://dx.doi.org/10.1038/nphoton.2012.360}{{\em Nature Photonics}
  {\bfseries 7}, 223--228 (2013)},
  \href{http://arxiv.org/abs/1110.4276}{{\ttfamily arXiv:1110.4276
  [quant-ph]}}.

\bibitem{shor08}
P.~W. Shor and S.~P. Jordan, ``Estimating Jones polynomials is a complete
  problem for one clean qubit,'' {\em Quantum Info. Comput.} {\bfseries 8},
  681--714 (2008), \href{http://arxiv.org/abs/0707.2831}{{\ttfamily
  arXiv:0707.2831 [quant-ph]}}.
  \url{http://dl.acm.org/citation.cfm?id=2017011.2017012}.

\bibitem{chiribella09}
G.~{Chiribella}, G.~M. {D'Ariano}, P.~{Perinotti}, and B.~{Valiron}, ``{Quantum
  computations without definite causal structure},''
  \href{http://dx.doi.org/10.1103/PhysRevA.88.022318}{{\em Phys. Rev.~A}
  {\bfseries 88}, 022318 (2013)},
  \href{http://arxiv.org/abs/0912.0195}{{\ttfamily arXiv:0912.0195
  [quant-ph]}}.

\bibitem{chiribella12}
G.~{Chiribella}, ``{Perfect discrimination of no-signalling channels via
  quantum superposition of causal structures},''
  \href{http://dx.doi.org/10.1103/PhysRevA.86.040301}{{\em Phys. Rev.~A}
  {\bfseries 86}, 040301 (2012)},
  \href{http://arxiv.org/abs/1109.5154}{{\ttfamily arXiv:1109.5154
  [quant-ph]}}.

\bibitem{colnaghi11}
T.~{Colnaghi}, G.~M. {D'Ariano}, S.~{Facchini}, and P.~{Perinotti}, ``{Quantum
  computation with programmable connections between gates},''
  \href{http://dx.doi.org/10.1016/j.physleta.2012.08.028}{{\em Phys. Lett.~A}
  {\bfseries 376}, 2940--2943 (2012)},
  \href{http://arxiv.org/abs/1109.5987}{{\ttfamily arXiv:1109.5987
  [quant-ph]}}.

\bibitem{araujo13b}
M.~Araújo, F.~Costa, and {\v C}.~Brukner, ``Decrease in query complexity for
  quantum computers with superposition of circuits,''.

\bibitem{oreshkov12}
O.~{Oreshkov}, F.~{Costa}, and {\v C}.~{Brukner}, ``{Quantum correlations with
  no causal order},'' \href{http://dx.doi.org/10.1038/ncomms2076}{{\em Nat.
  Commun.} {\bfseries 3}, (2012)},
  \href{http://arxiv.org/abs/1105.4464}{{\ttfamily arXiv:1105.4464
  [quant-ph]}}.

\bibitem{soeda13}
A.~Soeda, ``Limitations on quantum subroutine designing due to the linear
  structure of quantum operators,''. Talk at ICQIT2013.

\bibitem{thompson13}
J.~{Thompson}, M.~{Gu}, K.~{Modi}, and V.~{Vedral}, ``{Quantum Computing with
  black-box Subroutines},'' \href{http://arxiv.org/abs/1310.2927}{{\ttfamily
  arXiv:1310.2927 [quant-ph]}}.

\end{thebibliography}\endgroup
\end{document}